\documentstyle[epsfig,12pt]{article}

\begin{document}

\title{Strangeness production in proton-proton collisions
\thanks{Supported by Forschungszentrum J\"ulich,
the Jagellonian University of Cracow and the Cracow 
Institute of Nuclear Physics}
\thanks{Seminars held at Institute of Physics, Jagellonian
University.}}
\author{A. Sibirtsev and W. Cassing \\
Institut f\"ur Theoretische Physik, Universit\"at Giessen \\
D-35392 Giessen, Germany, \\
Institute of Physics, Jagellonian University \\
Reymonta 4, PL-30059 Cracow, Poland, \\
Institute of Nuclear Physics \\
Radzikowskiego 152, PL-31342 Cracow, Poland}
\date { }
\maketitle
\vspace{-9.5cm}
\hfill IFUJ, IFJ-1787/PH,\qquad  UGI-98-6
\vspace{9.5cm}
\newpage
\begin{abstract}
The cross sections of the reaction $pp \to N K Y$ for
$K^+$ or $K^0$ mesons and $\Lambda$ or $\Sigma$ hyperons
are calculated within the boson exchange model including 
pion and kaon exchange diagrams. We analyze the dependence 
of the results on the accuracy of the input $\pi N$ amplitude.
By fixing the $\pi NN $ coupling constant and the cut-off
parameter ${\Lambda}_{\pi}$ at the $\pi NN$ vertex 
we calculate the contribution from the kaon exchange 
diagram and obtain the ratio of the
$KN \Lambda $ and $KN \Sigma $ coupling constants by a
fit to the experimental data. This ratio is in a good agreement with
the $SU(6)$ prediction.  Our calculated total cross sections for the
different reaction channels are fitted by simple expressions 
and compared with other parameterizations used in the literature.
Furthermore, the gross features of the production cross section close
to threshold are discussed.
\end{abstract}
\newpage
\section{Introduction}
The production of strange particles in nuclear collisions is 
an exciting subject in nuclear physics; e.g. 
strangeness enhancement is proposed as a signature for the
formation of the quark qluon plasma in high energy 
nucleus-nucleus collision~\cite{Rafelski}. The 
production of strange particles 
in heavy-ion collisions at intermediate energies, furthermore, 
is also discussed as a way to study hot and dense 
nuclear matter~\cite{Brown,Cassing1} 
due to the weak $K^+N$ final state interaction.

During the last years  $K^+$ meson production from heavy-ion 
collisions at energies per nucleon below the free $NN$ 
reaction threshold was intensively studied at 
GSI~\cite{Grosse}, however, the interpretation
of the experimental data is strongly model dependent because secondary
production channels play a sensible role. 
Among the secondary reactions  relevant for strangeness
production at low energies are $\pi N$, $\Delta N$ and 
$\Delta \Delta$ collisions. As already discussed by Koch and 
Dover~\cite{Dover}  the $\Delta$  might play
an important role for subthreshold particle production. The
heavy-ion simulations from  Lang et al.~\cite{Lang}, 
Huang et al.~\cite{Huang}, 
Aichelin and Ko~\cite{Aichelin} and Li and Ko~\cite{Li} 
show that the GSI data might be reproduced when accounting for 
the kaon  production from the  $\Delta N$ and
$\Delta \Delta$ interactions. However, the former conclusion
strongly depends upon the elementary $NN \to NYK$
cross section employed and the ratios between $NN$, $\Delta N$
and $\Delta \Delta$ reaction channels. More recent calculations
on strangeness production in heavy-ion 
collisions~\cite{Cassing2}, which are based on more reliable 
elementary cross sections~\cite{Sibirtsev1}, indicate that the
$\pi N \to YK$ channels are dominant at low energies.
Note, that the strangeness
production via secondary meson-meson, meson-baryon and 
baryon-baryon interactions is also important 
for AGS energies \cite{Stachel}. 
Furthermore, experimental information on subthreshold 
strangeness production in  $pA \to K^+X$ reactions
is available from SATURNE~\cite{Senger} and
CELSIUS~\cite{Palmeri} and will be available at
COSY~\cite{Sistemich} soon. 

The most important ingredients for the 
proton-nucleus and heavy-ion simulations are the elementary
cross sections for strange particle production. From the
experimental data we know the $\pi N \to K Y$ amplitude
rather well whereas the $BB \to BYK$ cross section is not well
determined, since before 1997 the available experimental data on the
reaction  were very scarce~\cite{LB}.
A sizeable step forward was achieved recently
with the measurement of the $pp \to p \Lambda K^+$ cross section
by the COSY-11 Collaboration~\cite{Balewski}. This reaction was 
studied very close to threshold and could 
clarify the validity of the old parameterizations on strangeness
production from Randrup and Ko~\cite{Randrup1}
and Zwermann and Sch\"urmann~\cite{Zwermann}, which have  
peviously been  used for $p+A$ and $A+A$ simulations.

However, inspite of the experimental progress  our understanding
of the reaction mechanism is still poor.
The predictions of Li and Ko~\cite{LiKo} and 
Sibirtsev~\cite{Sibirtsev1} quite reasonably reproduce
the data~\cite{LB,Balewski} based on the 
$K$-meson exchange mechanism, while the study of
F\"aldt and Wilkin~\cite{Wilkin} and 
Tsushima et al.~\cite{Tsushima1,Sibirtsev2,Sibirtsev3}
indicate a dominance of $\pi$-meson exchange followed
by the excitation of  a $N^{\ast}(1650)$ baryonic resonance decaying
to $K Y$. In principle, by varying the model parameters both
approaches can describe the experimental data. Thus the most
constructive analysis of the present status
is to investigate the model uncertainties
as well as the selfconsistency of the calculations. 

Recently, the most detailed analysis in the resonance model
for strangeness production was performed by Sibirtsev, Tsushima and 
Thomas~\cite{Sibirtsev3}. Here the most crucial 
parameters entering the  $pp \to p \Lambda K^+$ 
calculations are efficiently fixed by the $\pi N \to \Lambda K$ 
reaction~\cite{Tsushima2} which clarifies the 
uncertainties of the resonance 
approach~\cite{Wilkin,Tsushima1}.
Here we analyze the role of the kaon exchange mechanism
for strangeness production and investigate to what
extent our results might be influenced by the model parameters.
Our study is organized as follows:
In Section 2 we review the theoretical status and the uncertainties
of the model while in Section 3 we formulate the 
one-meson exchange model.
Our results from the calculations with  pion and  kaon exchange 
diagrams are discussed in Sections 4 and 5, respectively.
In Section 6, furthermore, we analyze the 
near-threshold behavior of the
total production cross section while a summary is
presented in Section 7.

\section{Theoretical status and uncertainties}
\begin{figure}[h]
\epsfig{file=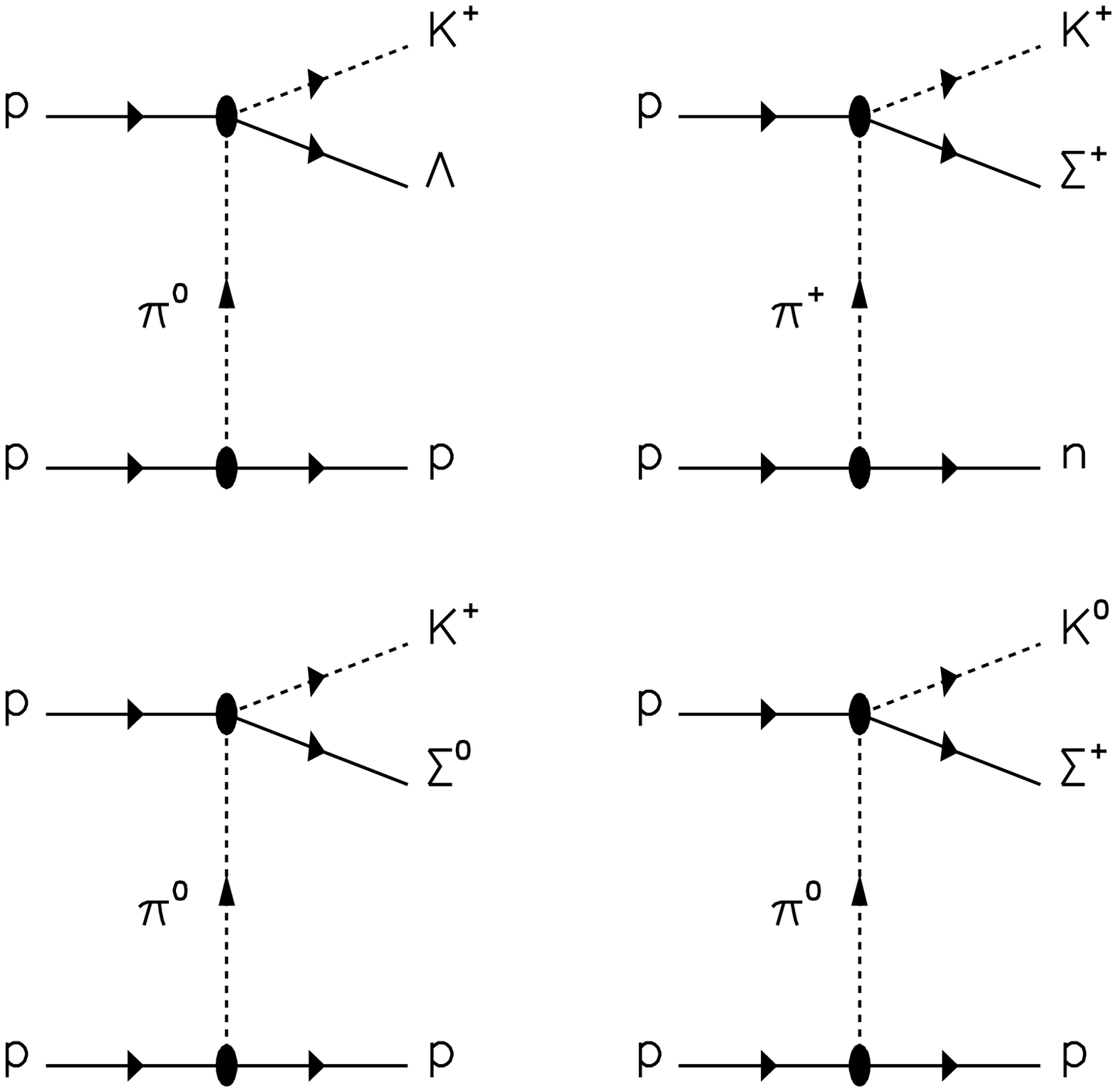,width=15cm}
\caption{\label{dia1}Pion exchange diagrams for  $pp \to pYK$
reaction channels.}
\end{figure}

\begin{figure}[h]
\epsfig{file=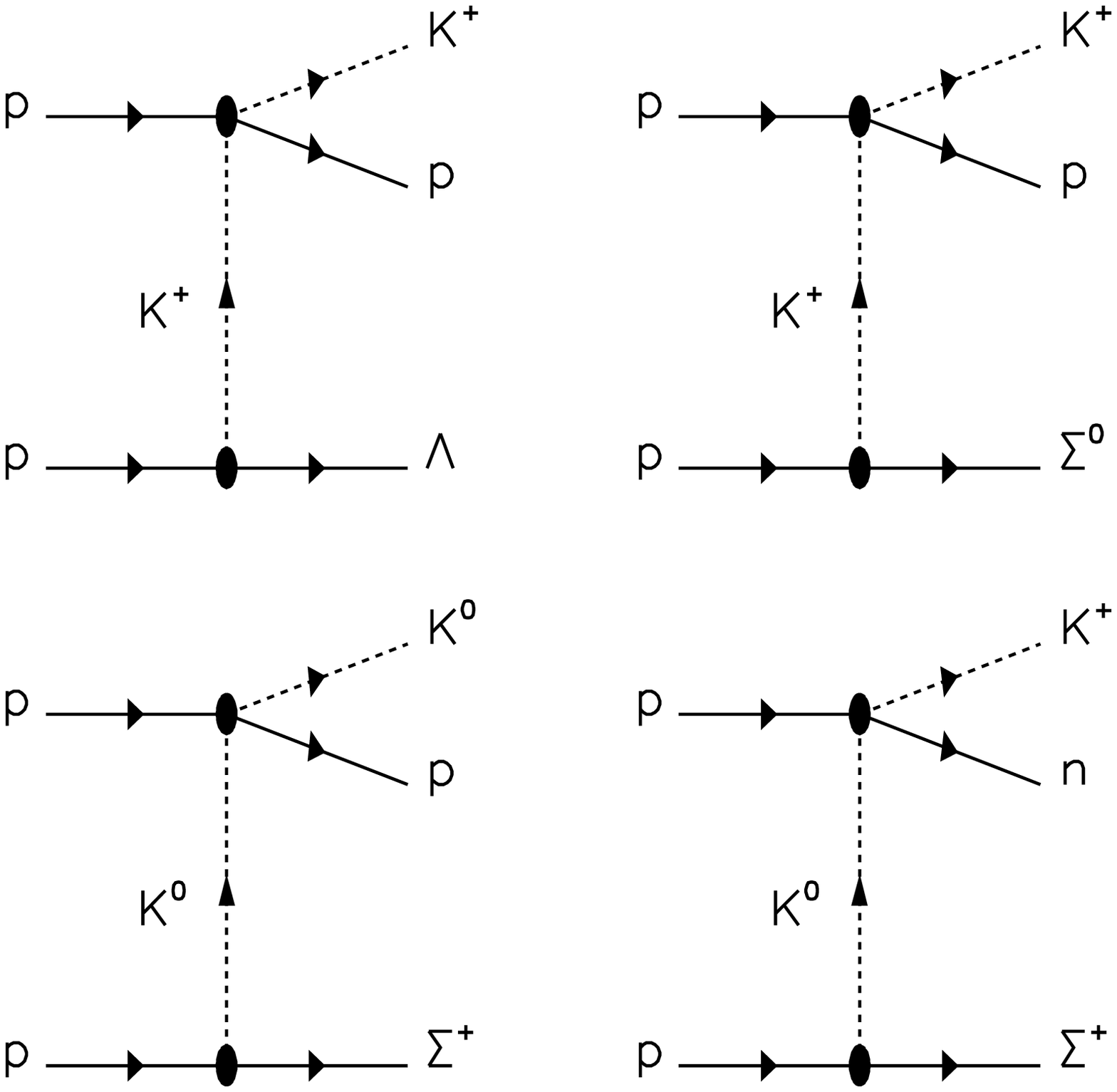,width=15cm}
\caption{\label{dia2}Kaon exchange diagrams for  $pp \to pYK$
reaction channels.}
\end{figure}

The forces between baryons are traditionally considered as due
to boson exchange. One of the models proposed to calculate the
nucleon-nucleon scattering is the  single
pion exchange~\cite{Ferrari1,Okun}.
An application of the meson exchange to
photoproduction was performed already in 1961 by 
Sakurai~\cite{Sakurai},
while for inelastic processes it was first investigated by
Chew and Low~\cite{Chew}.
A first application of the boson exchange mechanism to the 
calculation of strangeness production 
in proton-proton collisions at the energy of 3~GeV 
was performed in 1960 by Ferrari~\cite{Ferrari}.
Here, pion and kaon exchange graphs were considered
and it was found that the contribution from the pion exchange 
to the $pp \to p \Lambda K^+$ reaction cross
section is about twice less than that from
kaon exchange. The relevant diagrams for 
kaon production  from $pp$ collisions are shown in 
Figs.~\ref{dia1},\ref{dia2}.

Within a similar approach, but neglecting the contribution from 
the kaon exchange diagram, Yao~\cite{Yao} calculated the 
strangeness production in the 
reactions  $pp \to N Y K$ and $pp \to N Y K \pi $ at 2.9 GeV and
rather well reproduced the total cross section as well as momentum
and angular distributions. In his calculations it was assumed 
that the cut-off parameter of the pion form 
factor is given by the squared
nucleon mass while the energy dependence of the 
$\pi N \to Y K$ amplitude was neglected.
Lateron, Wu and Ko~\cite{Wu} calculated the energy dependence of 
the production cross section for the reaction $pp \to p \Lambda K^+$.
Using only the one pion exchange they could reasonably well 
fit the available experimental data and
thus pointed out a negligible contribution from the 
kaon exchange diagram.

A more detailed calculation of the
associated strangeness production in proton-proton collisions was
performed by Laget~\cite{Laget}. Within the boson
exchange model it was found that the pion exchange 
accounts only for about
20$\%$ of the total $pp \to p \Lambda K^+ $ cross section
whereas the residual cross section is due 
to the kaon exchange mechanism. 
It was also shown that the dominant contribution to the cross
section of the reaction $pp \to N \Sigma K $ stems from the
one-pion exchange diagrams due to the small coupling constant 
$g^2_{N \Sigma K} \simeq $0.8.

The most surprising results
were obtained by Deloff~\cite{Deloff} within the pion and 
the kaon exchange model. He found that the experimental data
can be reproduced by the pion and kaon exchange mechanism with the 
coupling constants $g^2_{N \Lambda K} \ll g^2_{N \Sigma K}$ which 
strongly contradicted
the previous calculations as well as the present understanding 
of the $YN$ interaction~\cite{Maessen,Reuber}.

More recent calculations on $BB \to NYK$ reactions
including proton, neutron and $\Delta$-resonances in the
entrance channel and accounting for both pion and
kaon exchanges were performed by Li and Ko~\cite{LiKo}
using the coupling constants and cut-off parameters from 
Ref.~\cite{Laget}. Moreover, similar to 
Laget~\cite{Laget} both the $\pi N \to Y K$ and $KN \to KN$
amplitudes were taken off-shell. The latter assumption is the
basic difference between our calculations and those from
Refs.~\cite{Laget,LiKo}. In our approach, to be described below,
we will adopt  on-shell amplitudes in the upper 
vertices in the diagrams
presented in Figs.~\ref{dia1},\ref{dia2}. 

\begin{figure}[h]
\epsfig{file=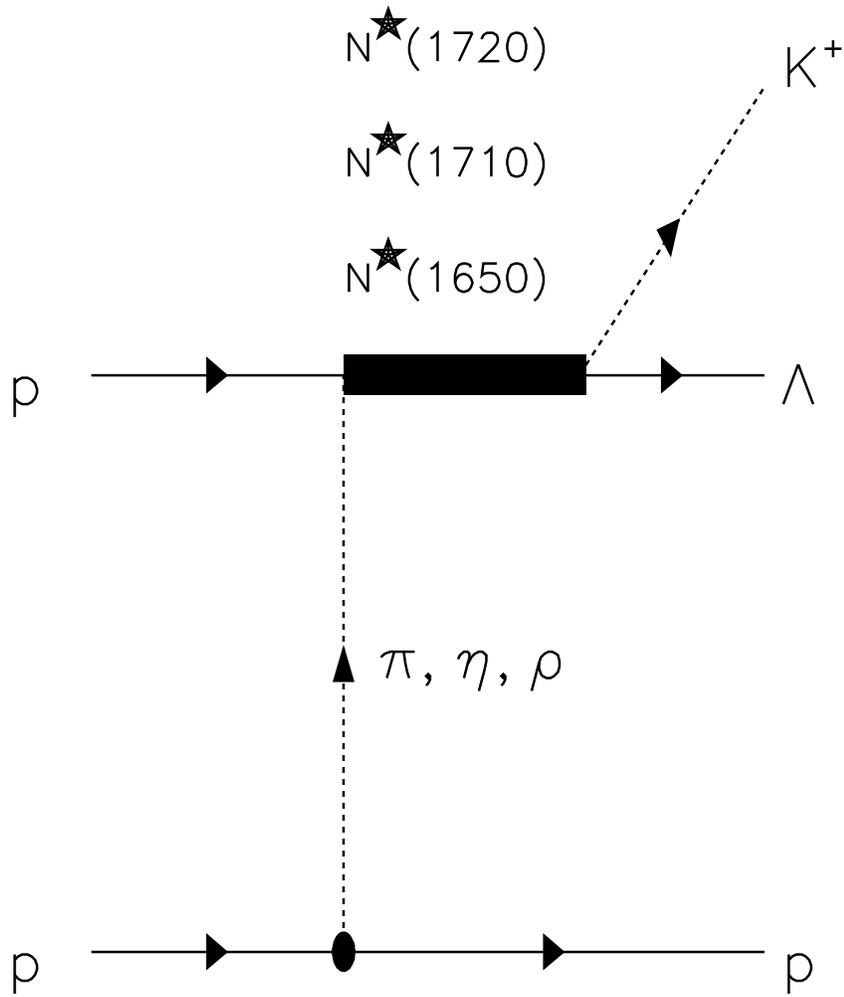,width=15cm}
\caption{\label{dia3}Meson exchange diagram for  
$pp \to p\Lambda K^+$ reaction in terms of the resonance model.}
\end{figure}

Calculations of the $K$-meson production
in proton-proton collisions close to threshold within 
the one-pion exchange 
followed by $N^{\ast}(1650)$ resonance excitation were
performed by F\"aldt and Wilkin~\cite{Wilkin}. Actually
the pion exchange diagrams from Fig.~\ref{dia2} can be
reversed to the graph shown in Fig.~\ref{dia3}. Moreover,
within this resonance model one has to account for all
baryonic resonances $R$ coupling to $R \to YK$ as well as to
$R\to \pi N$, $R \to \eta N$ and $R \to \rho N$ channels. Thus the
resonance model allows to treat not only the pion exchange,
but also the $\eta $ and $\rho$-meson exchanges simultaneously.
Since the $N^{\ast}(1650)$ is the lowest baryonic resonances,
which can decay to the $\Lambda K$ channel,  its contribution
is dominant at low collision energies. Since
$N^{\ast}(1650)$ is not coupled to the $\Sigma $ channel, one
has to account for other baryonic resonances
to calculate $pp \to N \Sigma K$ cross sections~\cite{Tsushima1}.

The next $N^{\ast}(1710)$ resonance coupled to the $\Lambda K$ channel 
has a larger mass and can be excited in $pp$ collisions at 
energies around 100~MeV above the $pp \to p \Lambda K^+$
reaction threshold. Note that the $N^{\ast}(1650)$ does not
decay to the $\Sigma K$ channel and for the
$pp \to N \Sigma K$ reaction it is necessary to incorporate
both the $N^{\ast}(1710)$ and $\Delta (1920)$ resonances.
Since the $N^{\ast}(1650)$ strongly couples to the pion (60-80\%)
and only slightly to the $\eta$-meson (3-10\%), the
motivation of F\"aldt and Wilkin~\cite{Wilkin} appeared quite
reasonable. The actual question is the absolute value of
their prediction for the $pp \to p \Lambda K^+$ cross section.

The $N^{\ast}(1650) \to N \pi$ and 
$N^{\ast}(1650) \to \Lambda K$ branching ratios
are known experimentally~\cite{PDG} and therefore the 
corresponding coupling constants  can be reasonably 
fixed on resonance. 
Note that by changing the couplings one can vary only the absolute 
magnitude of the 
$pp \to p \Lambda K^+$ cross section, but not its energy dependence.
Thus, in principle, keeping these constants as free parameters
one can perfectly reproduce the experimental data
near threshold~\cite{Balewski}. However, these
parameters can be varied only within a certain range in line with
the particle properties~\cite{PDG}. 
A more consistent calculation within the resonance model and for the
$pp \to N \Lambda K$ and $pp \to N \Sigma K$ reactions
was performed by Tsushima, Sibirtsev and 
Thomas~\cite{Tsushima1,Sibirtsev3}.
To avoid the uncertainties in the coupling constants and cut-off
parameters the latter were taken from the analysis 
of the $\pi N \to \Lambda K$
and $\pi N \to \Sigma K$ reactions. Moreover, their calculations
were performed for a wide range of $pp$ collision energies
in order to reproduce the recent near-threshold experimental
data~\cite{Balewski} as well as the results available
at high energies~\cite{LB}. All available and relevant 
baryonic resonances 
as well as $\pi$, $\eta$ and $\rho$-meson exchanges were
properly included. However, it was found that 
the calculations within the 
resonance model~\cite{Tsushima1,Sibirtsev2} underestimate the
$pp \to p \Lambda K^+$ cross section at low energy.

Let us now discuss the uncertainties of the calculations
within the pion and kaon exchange model.
Obviously, the calculations with the boson exchange model 
following the prescription from Figs.\ref{dia1},\ref{dia2}
substantially depend on the accuracy of the  
$\pi N \to YK$  and $KN \to KN$ amplitudes entering the calculations. 
Again, quite reliable studies for the 
$\pi N \to YK$ reaction within the resonance 
model were done by Tsushima, 
Huang and Faessler~\cite{Tsushima2} who perfectly 
describe all channels of the $\pi N \to YK$ reaction at low 
energies. The $\pi N \to YK$ reaction was also investigated 
within the microscopic quark model by Yan, Huang and
Faessler~\cite{Yan}. Moreover, the experimental data on
strangeness production in pion induced reactions are 
well suited to perform simple 
parametrizations~\cite{Cugnon1,Cugnon2} of the
$\pi N \to YK$ cross sections and to adopt them for the
$pp \to NYK$ calculations within the pion-exchange model.

On the other hand, the  $KN \to KN$ amplitude cannot be well
determined from the experimental data, since these
are available for the $I=1$ channel only, which corresponds 
to the reaction $K^+ p \to K^+ p$. The reconstruction of the
$I=0$ amplitude is not very accurate~\cite{Martin,Giacomelli},
but this cross section is important for the kaon
exchange model because it contributes to the following 
two-body amplitudes:
\begin{eqnarray}
M(K^+p \to K^+p) = M(K^0n \to
K^0n)=M_1 \nonumber \\
M(K^+n \to K^+n) =M(K^0p \to
K^0p) = \frac{1} {2}  \ (M_0+M_1) \\
M(K^+n \to K^0p)=M(K^0p \to
K^+n) = \frac {1} {2}  \ (M_1-M_0)
\nonumber
\end{eqnarray}
A substantial progress in the calculation of the
$M_0$ amplitude was achieved recently by the J\"ulich 
group~\cite{Hoffman} in analyzing $KN$ elastic 
cross sections. On the other hand, to calculate the
contribution from the kaon exchange to the
$pp \to p \Lambda K^+$ cross section, one needs the $I=1$
amplitude only. The coupling constants 
$g_{NN \pi}$, $g_{N \Lambda K} $ and
$g_{N \Sigma K} $ as well as the corresponding cut-off
parameters substantially influence  
the relative contribution from the pion and 
the kaon exchange mechanism.  
Actually we consider the coupling constant $g_{NN \pi}$ as well as the
cut-off parameter ${\Lambda}_{\pi}$ in the $NN \pi$ vertex 
to be fixed by the analysis in the Bonn model~\cite{Machleidt1}.

Furthermore, the  $g_{N \Lambda K} $ and $g_{N \Sigma K}$ couplings
are related to the mixing angle $\epsilon $ between the $\Lambda $ and 
$\Sigma$ contents of the nucleon as
\begin{equation}
{\cal R } =\frac {g^2_{KN \Lambda}} {g^2_{KN \Sigma}} = 3  \
cot^2 \epsilon .
\end{equation}
$SU(6)$ symmetry  predicts $\epsilon =18.4^0 $  and thus a ratio 
${\cal R}=g^2_{N \Lambda K}/ g^2_{N \Sigma K} \approx$
27/1.  Note that when using the resonance prescription the 
coupling constant is related to the partial decay width of the
resonance and thus the $g_{NYK }$ coupling defines the
strangeness content $N \to Y K$ of the nucleon. This clearly
indicates the importance of the $NN \to NYK$ reactions for the 
understanding of the fundamental properties of the nucleon. 

In Table 1 we summarize the results on ${\cal R }$ 
obtained from the $KN$ dispersion relations~\cite{Martin1,Dalitz}.
The coupling constants from Dalitz et al.~\cite{Dalitz} 
were obtained by an analysis of the total set of experimental 
data on $KN$ scattering. 
The result from the meson exchange calculation of  
Laget~\cite{Laget} is close to the prediction from $SU(6)$ symmetry,
while the ratio ${\cal R}$ from Deloff~\cite{Deloff}
is the smallest value among those obtained presently. 
The result of Siebert et al.~\cite{Siebert} was adjusted to the 
experimental data from SPES4  at SATURNE.

\begin{table}[h]
\begin{center}
\caption{Cut-off parameters and coupling constants from
different studies}
\vspace{0.5cm}
\begin{tabular}{|l|c|c|c|c|c|}
\hline
Reference & ${\Lambda}_K$ [GeV] & ${\Lambda}_{\pi}$ [GeV] & 
$g^2_{N \Lambda K} $ & $g^2_{N \Sigma K} $ & ${\cal R}$  \\
\hline
Martin \protect\cite{Martin1} & & & $13.9 \pm 2.6$ & $ 3.3 \pm 1$ 
& $4.2 \pm 1.5$ \\ 
McGinley \protect\cite{Dalitz} & & & 9.0 & 1.5 & 6  \\
Dalitz \protect\cite{Dalitz} & & & 20.7 & 1.0 & 20.7 \\
Laget \protect\cite{Laget} & 0.85 & 1.2 & 14 & 1 & 14 \\
Deloff \protect\cite{Deloff} & & & & & 0.08 \\
Siebert \protect\cite{Siebert} & & & & & 1.6 \\
Our results & $0.81 \pm 0.14$ & 1.0 & $19.6 \pm 4.2$ 
& $1.3 \pm 0.3 $ & $15 \pm 6 $ \\
$SU(6)$ & & & & & 27 \\   
\hline 
\end{tabular}
\end{center}
\end{table}

We note, that the interference due to the exchange of the 
nucleons for the $s$-wave phase shift, 
which is relevant for
$K^+p \to K^+p $ elastic scattering at low energies, might 
change the model results by around 30$\%$~\cite{Ferrari3}.
However, this interference is neglected in most of the calculations.
Furthermore, the interference  term between pion and kaon exchange 
substantially depends on the phase shifts of the
$\pi N \to Y K$ and $KN \to KN$ amplitudes.
To fix the relative sign  between the $\pi N \to Y K$ and 
$K N \to K N$ exchange amplitudes one needs a well-defined
interacting Lagrangian~\cite{LiKo}. Alternatively, Laget 
proposed~\cite{Laget}  to choose the sign in order to maximize the 
strangeness production cross section.

The analysis of the differential cross sections for the
reaction $pp \to N Y K$ were performed  in 
Refs.~\cite{Laget,Deloff,Siebert} and contain much more
uncertainties. To calculate the kaon  as well as the baryon 
spectra one needs to incorporate the $t$-dependence of the 
$\pi N \to Y K$ and $KN \to KN $ amplitudes. 
It is also discussed that the  hyperon-nucleon interaction in
the final state might  change the observed spectra. 
In this sense the analysis of the total
production cross section contains less uncertainties. 

Actually there is no straightforward way to perform the
calculations with such large uncertainties in the
model parameters as the coupling constants, cut-offs as well
as $\pi N \to Y K$ and $K N \to K N$ amplitudes.
Therefore, in order to proceed the calculation 
we propose the following scheme:

We assume that the coupling constant $g_{NN \pi}$ and the cut-off
parameter ${\Lambda}_{\pi}$ are fixed by
the Bonn model~\cite{Machleidt1} as well as 
an analysis of pion production from proton-proton
collisions~\cite{Machleidt1,Kaidalov1} and
first calculate the cross section of the reactions
$pp \to p \Lambda K^+$ and $pp \to N \Sigma K$ using 
pion exchange, only.
The difference between the experimental data and the 
one-pion exchange model calculations
then is fitted by the kaon exchange mechanism in order to obtain
the ratio $\cal R $ and the cut-off parameter ${\Lambda}_K$
neglecting the interference terms and assuming isotropy of the
$KN$ elastic cross section at energies near the 
kaon production threshold.

\section{The one-boson exchange formalism}
\begin{figure}[h]
\epsfig{file=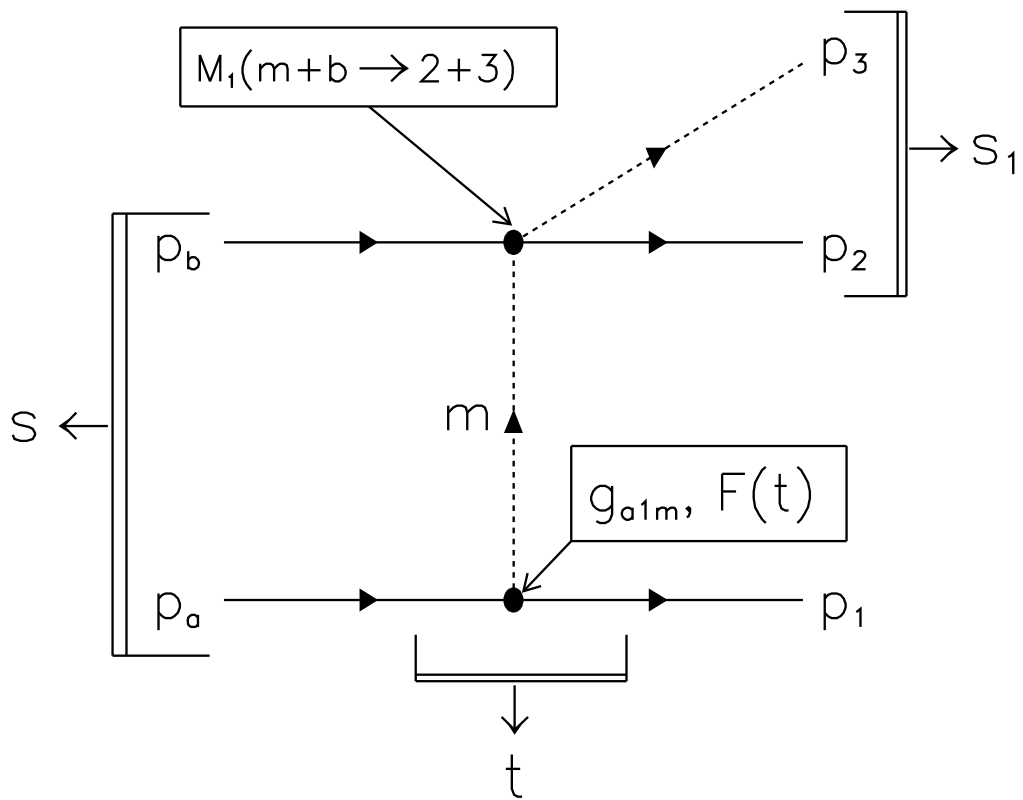,width=14cm}
\caption{\label{dia0}Feynman diagram for  
$a+b\to 1+2+3$ reaction.}
\end{figure}

A detailed description of the one-boson exchange model
is given by Ferrari and Selleri~\cite{Ferrari1} and 
Berestetsky and Pomeranchuk~\cite{Pom}. Consider a physical
reaction in which two initial particles with 4-momenta $p_a$
and $p_b$ transform into three particles,
\begin{equation}
\label{react}
p_a+p_b \to p_1+p_2+p_3
\end{equation}
with final 4-momenta $p_1$, $p_2$ and
$p_3$. Fig.~\ref{dia0} illustrates the Feynmann diagram of
the process. 

Taking the particle $a$ of spin 1/2 and the 
exchange of a single spineless boson $m$ with mass $\mu$,
the amplitude for this reaction $M$ is given within the 
pole approximation as
\begin{equation}
\label{form1}
M= g_{a1m} \ F(t)\  {\bar u(p_1)}\ O \ u(p_a)
\ \frac{1}{t-{\mu}^2} M_1 ,
\end{equation}
where $g_{a1m}$ is the coupling constant of the $(a \to 1+m)$
vertex, $F(t)$ is the form factor of the proper vertex part,
$O$ is the operator (which is ${\gamma}_5$ or 1 according 
to the parity  of the exchanged boson) and $t=(p_a-p_1)^2$ is
the 4-momentum transfer squared. 
In Eq.~\ref{form1} $M_1$ is the amplitude of 
the process $m+b \to 2+3$, which is related to the physical
cross section as~\cite{Byckling}
\begin{equation}
\label{form2}
|M_1|^2 = 64  \ \pi^2 \ s_1 \frac{q_m}{q_2} \ 
\frac{d\sigma (m+b\to 2+3)}{d\Omega},
\end{equation}
where $s_1=(p_2+p_3)^2$ is the squared 
invariant mass of the (2+3) system,
while $q_m$ and $q_2$ are the momenta of the corresponding particles 
in the center-of-mass for this system. In this way one is allowed to
introduce experimental information in the calculations within
the boson exchange model.

The pole approximation implies that the formalism
is valid within the limit $|t| \to \mu^2$ with $\mu$ being the
mass of the exchanged particle. As was discussed
in Ref.~\cite{Pom}, at small values of the transfered momentum,
$|t|\le \mu^2$,
the pole term~(\ref{form1}) should yield the correct magnitude of
the cross section~(\ref{react}) while at large $|t|$ 
one should take care about the corrections to the amplitude $M$.

Now the double differential cross section for the 
reaction~(\ref{react}) can be expressed as
\begin{equation}
\label{form4}
\frac{d^2\sigma}{dt ds_1} = \frac{1}{2^9 \pi^3 q_a^2 s}
\ \frac{q_2}{\sqrt{s_1}} \ |M|^2 ,
\end{equation}
where $s=(p_a+p_b)^2$ is the squared invariant mass of the
colliding particles and $q_a$ is the momentum of particle $a$ in
their center-of-mass. Substituting the squared amplitude from
Eq.~\ref{form1} one can easily obtain the general expression
\begin{equation}
\label{form3}
\frac{d^2\sigma}{dt ds_1} = \frac{g^2_{a1m}}{32\pi^2 q_a^2s}
\ q_m  \ \sqrt{s_1} \frac{F^2(t)}{(t-\mu^2)^2} 
\ \left[ (m_1 \pm m_a)^2-t \right] \ {\sigma}(m+b \to 2+3),
\end{equation}
where a plus sign has to be chosen if the exchanged meson is
scalar and a minus sign if it is pseudoscalar~\cite{Bjorken}.
The spin summation and averages of $|M|^2$ are assumed to be included
and not written explicitly. 
Besides the  form factor $F(t)$ Eq.~\ref{form3} is the
same as in Ref.~\cite{Ferrari3} performed within the $S$-matrix
approach. 

The total cross section can now be obtained by integrating the
expression~(\ref{form3}) over the momentum transfer and
the invariant mass of the (2+3) system. The ranges of integrations
are given in Ref.~\cite{Byckling}.

A form factor is introduced in order to avoid the
divergence of the total cross section at large collision
energy or large momentum transfer.
The sensitivity of the model results due to different types of 
form factors were studied in detail in the J\"ulich-Bonn 
potential model~\cite{Hoffman,Machleidt1,Juelich}. We note that
one of the relevant historical methods is to cut the  integration of 
Eq.~\ref{form3} by a  proper choice of the maximal value of $|t|$.
Another well known method is to use a 'reggeized' 
boson exchange~\cite{Regge}.

\section{The pion exchange model}
In our  calculations we account for the one-pion exchange diagrams 
shown in Fig.~\ref{dia1}. Within the  pion exchange model 
the cross sections for the reactions
$pp \to p \Lambda K^+$ and $pp \to N \Sigma K$ are given as 
\begin{eqnarray}
\label{OME}
\sigma (pp \to NYK, \sqrt{s}) = \frac {m_N^2} {2 {\pi}^2 q_i^2 s}
\int _{W_{min}}^{W_{max}} k \  W^2 \ \sigma (\pi N \to K Y, W) \
 dW \times \nonumber \\
\int _{t_-}^{t^+} \frac {f^2_{NN \pi} } {{\mu}^2} \ F^2(t) 
\ D^2(t) \ t \ dt ,
\end{eqnarray}
where $\sqrt{s}$ and $W$ are the invariant masses of the colliding
protons and the produced kaon-hyperon system, respectively. Obviously
we have
\begin{equation}
W_{min}=m_Y +m_K \ \ \ and \ \ \ W_{max}= \sqrt{s}-m_N
\end{equation}
with $m_N$, $m_Y$ and $m_K$ being the masses of the 
nucleon, $\Lambda $ or $\Sigma $-hyperon and $K$-meson,
respectively.
In Eq.~(\ref{OME}) $t$ stands for the squared four-momentum
transfer from the initial to the final nucleon and
\begin{equation}
t_{\pm} = 2m_N^2-2E_iE_f \pm 2q_iq_f   ,
\end{equation}
where $E_i$, $q_i$ are the energy and the momentum of the initial
nucleons in the center-of-mass frame, while $E_f$, $q_f$ are that 
for the final nucleons; $\mu$ and $k$ denote the mass and momentum
of the exchange pion. With the kinematical function
\begin{equation}
\lambda (x, y, z) = (x-y-z)^2-4yz 
\end{equation}
one can simply express the momenta as
\begin{eqnarray}
q_i^2 & = & {\lambda}(s, m_N^2,m_N^2)/4s \nonumber \\
q_f^2 & = & {\lambda}(s, W^2, m_N^2)/4s  \\
k^2 & = & {\lambda}(W^2, m_N^2, {\mu}^2)/4W^2 . \nonumber
\end{eqnarray}  

The pion propagator is given by
\begin{equation}
D(t)= \frac {1} {t-{\mu}^2};
\end{equation}
the coupling constant  $f^2_{NN \pi}=$1.0  is similar to
that used in Refs.~\cite{Germond,Laget,Vetter,LiKo}.
Note, that the renormalized coupling constant $f_{NN \pi}$
is related to $g_{NN \pi}$ as~\cite{Pom}
\begin{equation}
\frac{f^2}{4\pi} = \frac{g^2}{4\pi} \ 
{\left(\frac{\mu}{2m_N} \right)}^2
\end{equation}
with ${g^2}/{4\pi}$ = 14.4~\cite{Machleidt1}. 
To account for the off-shell modification of
the $NN \pi $ vertex we use a monopole  form factor 
\begin{equation}
\label{formf}
F(t) = \frac {{\Lambda}_{\pi}^2 - {\mu}^2} 
{ {\Lambda}_{\pi}^2 - t}
\end{equation}
with the pion cut-off parameter ${\Lambda}_{\pi} =$1~GeV.
Furthermore, $\sigma (\pi N \to K Y, W)$ is the 
cross section for the corresponding  $\pi p \to K Y$ channel. 
Isospin symmetry is applied in order to get the relations
between the different channels with $\Lambda$ 
and $\Sigma $ hyperon production, i.e.
\begin{eqnarray}
\sigma ({\pi}^0 p \to K^+ \Lambda) & = &  \frac {1} {2} 
\sigma ({\pi}^- p \to K^0 \Lambda) \nonumber \\
\sigma ({\pi}^0 p \to K^0 {\Sigma}^+) & = &
\sigma ({\pi}^+ n \to K^+ {\Sigma}^0) . 
\end{eqnarray}

As already discussed in Section~2 the $\pi N \to K Y$
cross section may be taken either from the calculations 
within the resonance model~\cite{Tsushima2} or in form of a
parameterization of the relevant experimental data~\cite{Cugnon1}. 

Let us to remind that the difference of the present model 
prescription of the $pp \to NYK$ reaction to that 
proposed by Li and Ko~\cite{LiKo} is due to the introduction 
of the  form 
factor~(\ref{formf}) in the upper $\pi N Y K$ vertex of the
diagrams in Fig.~\ref{dia1}  
accounting for the off-shell nature of the exchanged pion.
This correction was included in the calculations~\cite{LiKo}
and thus the results from~\cite{LiKo} should actually be 
lower than our calculations.
We do not correct the $\pi N \to Y K$ amplitude since in the
calculations within the resonance model~\cite{Tsushima2} 
the corresponding 
form factor was already introduced. Thus using the 
parameterization from~\cite{Tsushima2} we effectively already 
account for the proper cross section.

The cross sections for the different channels of the
reaction $\pi N \to Y K$ can be parameterized as
~\cite{Tsushima2}  
\begin{equation}
\label{PART}
\sigma (\pi N \to K Y) = \sum_{j}
\frac {A_j (\sqrt{s} -\sqrt{s_{th}} )^{f_j}}
{(\sqrt{s} -M_j)^2 + B_j^2},
\end{equation}
where $\sqrt{s}$ is the invariant mass of the $\pi N$ system,
$\sqrt{s_{th}}=m_K+m_Y$  was taken to be equal 1.613 GeV
for $\Lambda K$ and 1.688 GeV for $\Sigma K$ production. The parameters
$A_j$, $f_j$, $M_j$ and $B_j$ are listed in Table~2 for the 
relevant reaction channels. 

\begin{table}[h]
\begin{center}
\caption{Parameters of the approximation (\protect\ref{PART}).}
\vspace{0.5cm}
\begin{tabular}{|l|c|c|c|c|c|}
\hline
Reaction & $j$ & $A$ [$\mu $b] & $f$ & $M$ [GeV] & $B$ [MeV] \\
\hline
${\pi }^-p \to K^0 \Lambda $ & 1 & 7.665 & 0.1341 & 1.72 & 88.465 \\
${\pi }^+p \to K^+ {\Sigma}^+ $ & 1 & 35.91 & 0.9541 & 1.89 &
124.418 \\
${\pi }^+p \to K^+ {\Sigma}^+ $ & 2 & 159.4 & 0.01056 & 3.0 
& 970.155 \\
${\pi }^+n \to K^+ {\Sigma}^0 $ & 1 & 50.14 & 1.2878 & 1.73 & 
80.343 \\
${\pi }^0p \to K^+ {\Sigma}^0 $ & 1 & 3.978 & 0.5848 & 1.74 &
81.67 \\
${\pi }^0p \to K^+ {\Sigma}^0 $ & 2 & 47.09 & 2.165 & 1.905 &
79.737 \\ 
\hline 
\end{tabular}
\end{center}
\end{table}

In order to study the sensitivity of the results within the
pion exchange model on the accuracy of the incorporated $\pi N$
amplitude we also test the reaction $p p \to p \Lambda K^+ $
with two other parameterizations from 
Cugnon et al.~\cite{Cugnon1,Cugnon2}, i.e.: 
\begin{eqnarray}
\label{PARC1}
\sigma ( {\pi}^-p \to K^0 \Lambda) & = &
9.8 (\sqrt{s} -1.6) \ \   at  \ \ \sqrt{s} <1.7 \ GeV
\nonumber \\
& = & 0.084 (\sqrt{s} - 1.6)^{-1} \ at  \ \ 
\sqrt{s} \ge 1.7 \ GeV
\end{eqnarray}
where the cross section is given in $mb$ and
\begin{eqnarray}
\label{PARC2}
\sigma ({\pi}^-p \to K^0 \Lambda ) & = & 0.65 p^{4.2}
\ \  at  \ \ 0.9<p<1.0 \nonumber \\
& = & 0.65 p^{-1.67} \ \  at  \ \ p \ge 1 
\end{eqnarray}
with the cross section given again in $mb$ and the incident pion 
momentum in the laboratory frame $p$ taken in GeV/c.

\begin{figure}[hb]
\epsfig{file=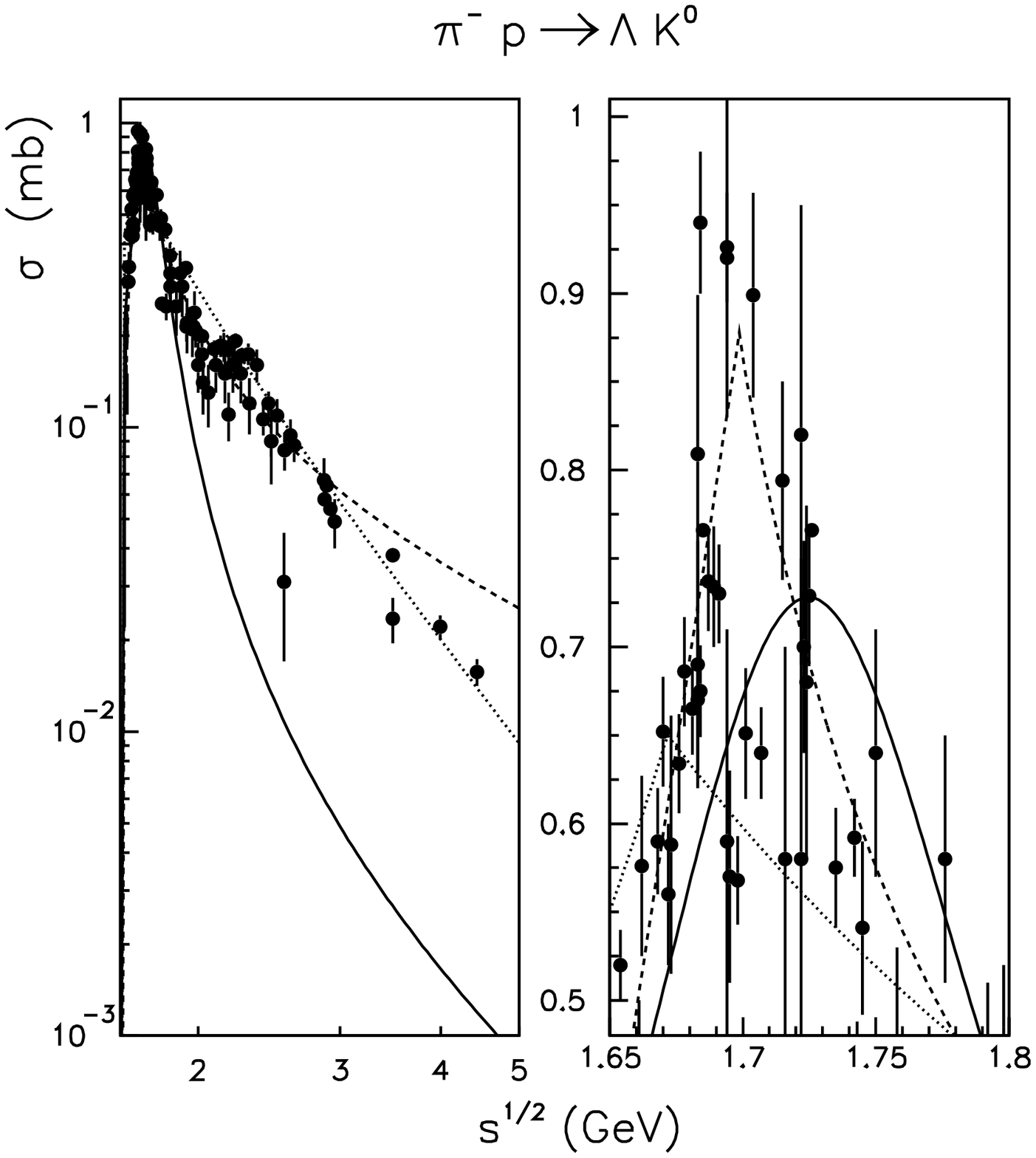,width=15cm}
\caption{\label{ka2}The cross section for the reaction
${\pi}^-p \to K^0 \Lambda$. Experimental points 
are from~\protect\cite{LB}. The lines show the parameterizations 
\protect\cite{Tsushima2}-solid, \protect\cite{Cugnon1}-dashed
and \protect\cite{Cugnon2}-dotted}
\end{figure}

Fig.~\ref{ka2} shows the experimental cross section of the reaction
${\pi}^-p \to K^0 \Lambda$ from  Ref.~\cite{LB} together with the 
parameterizations~(\ref{PART}) and~(\ref{PARC1}). We stress 
the difference between the parameterizations
at $\sqrt{s}>2$ GeV which influences the results from
the pion exchange model at large collision energies. However, in the
maximum of the ${\pi}^-p \to K^0 \Lambda $ cross section
the difference between (\ref{PART}) and (\ref{PARC1}) is only 
about 30$\%$. The dotted line in Fig.~\ref{ka2}
indicates the  parameterization~(\ref{PARC2}). 

\begin{figure}[hb]
\epsfig{file=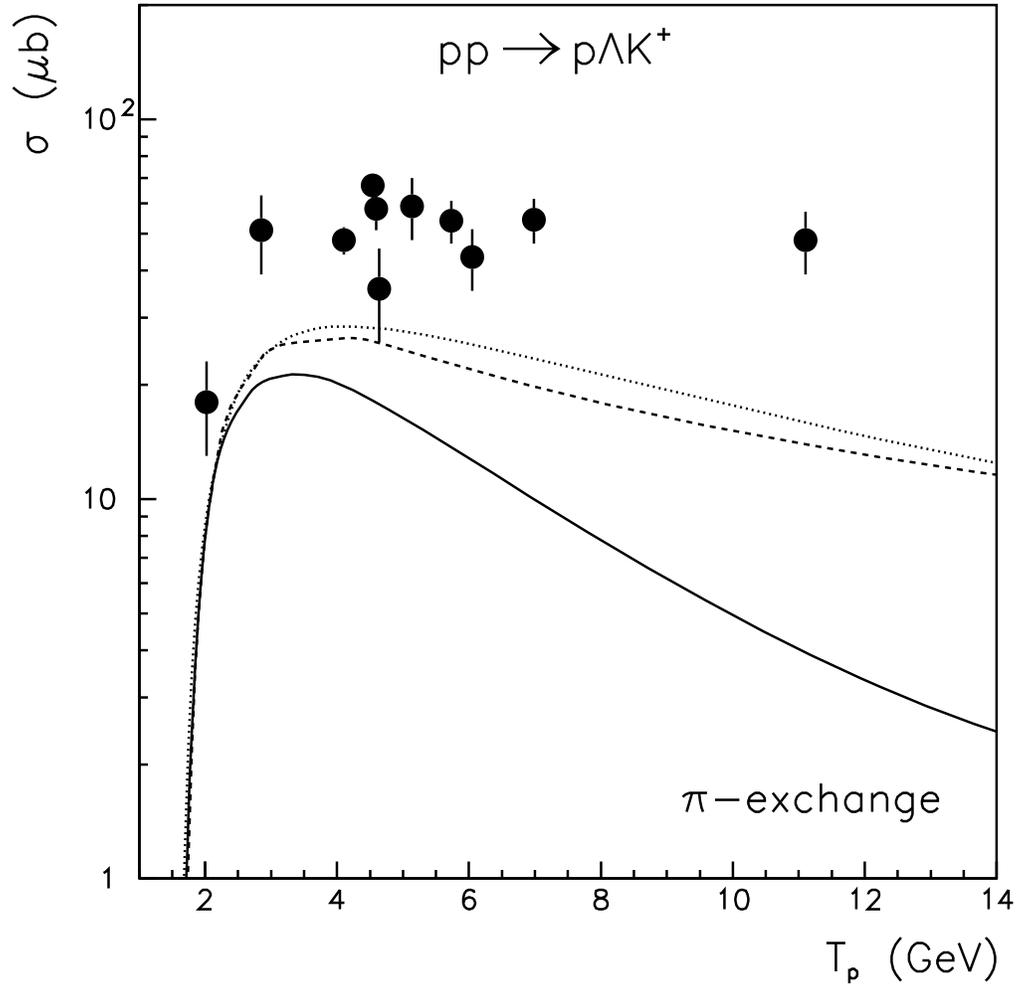,width=15cm}
\caption{\label{ka1}The cross section for the reaction
$pp \to p \Lambda K^+$. Experimental points
are from \protect\cite{LB}. The lines show the results from 
the pion exchange model calculated with
different  $\pi N$ amplitudes. The notations are the same as in  
Fig.~\protect\ref{ka2}.}
\end{figure}

The resulting cross section for the reaction $pp \to p \Lambda K^+$
within the one-pion exchange model for 
different parameterizations of $\sigma ({\pi}^0 p \to
K^+ \Lambda )$ together with the experimental
data~\cite{LB} are shown in Fig.~\ref{ka1} as
a function of the proton beam energy.  The notations  are
similar to those in Fig.~\ref{ka2}. 

We conclude that, i) the results from the pion exchange
calculations with all different $\pi N \to \Lambda K$
parameterizations underestimate the experimental data 
for the reaction $pp \to p \Lambda K^+$ 
by a factor of 3. This discrepancy indicates the importance
of the kaon exchange mechanism; 
ii) we find that the results strongly depend on the
input $\pi N \to K \Lambda$ amplitude at high energies. 
Indeed, this was expected since calculations from 
Ref.~\cite{Tsushima2} did not reproduce the 
$\pi N \to \Lambda K$ cross section at high energies
(cf. Fig.~\ref{ka2}). On the other hand, at low energies 
there are no differences 
between the results obtained with different 
$\pi N \to \Lambda K$ parameterizations. 
Moreover, the accuracy of the experimental data for $pp$ collisions is
better than the differences from the different
$\pi N \to \Lambda K$ parameterizations. 
In the following calculations we thus use ~(\ref{PART}).

\begin{figure}[h]
\epsfig{file=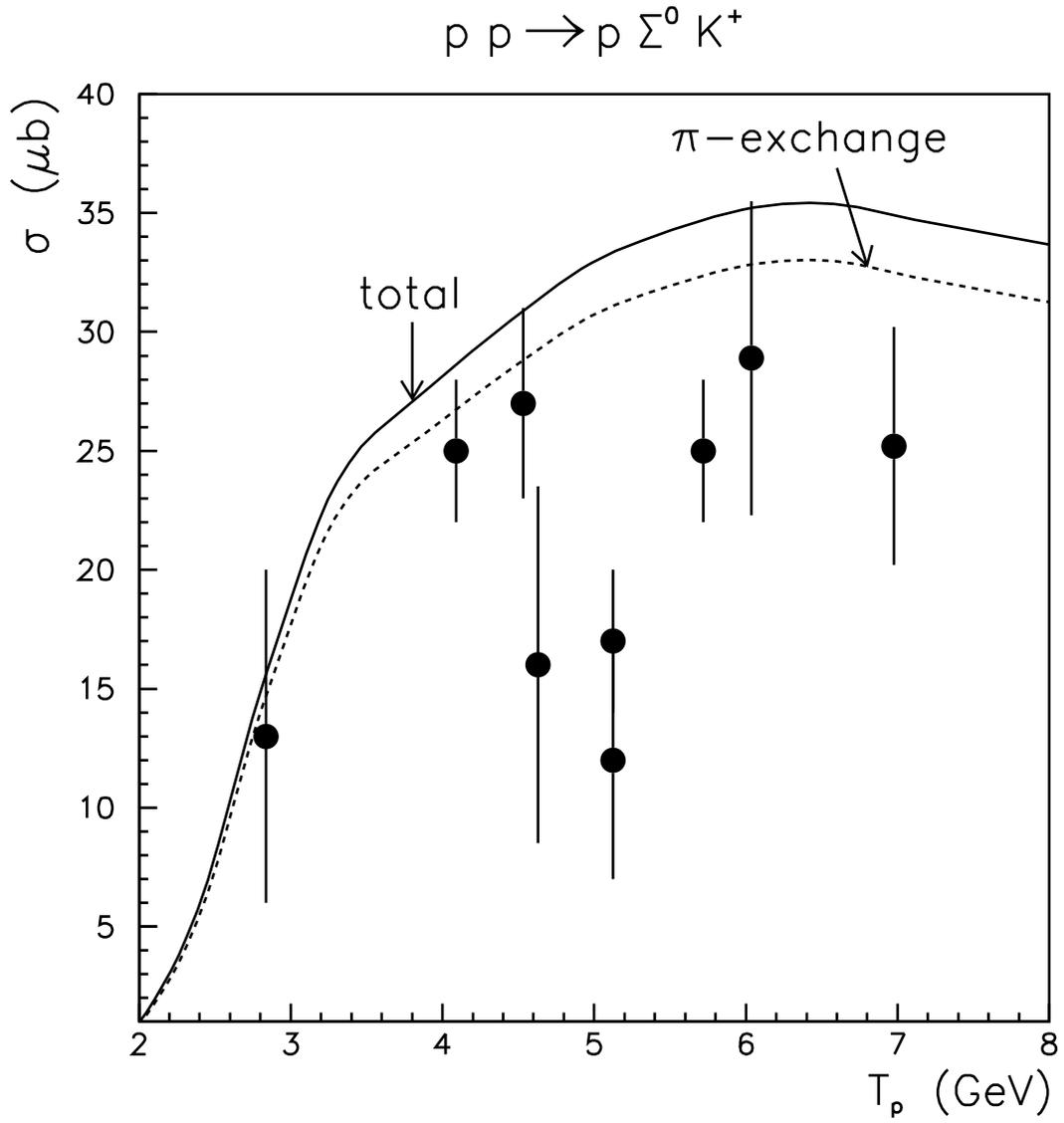,width=16cm}
\caption{\label{ka5-1}The cross section for the reaction
$pp \to p \Sigma^0 K^+$. Experimental points
are from~\protect\cite{LB}. The lines show the results from 
the boson exchange model. The solid line is the sum of the
pion and kaon exchange; the dashed line shows pion exchange, only.}
\end{figure}

\begin{figure}[h]	
\epsfig{file=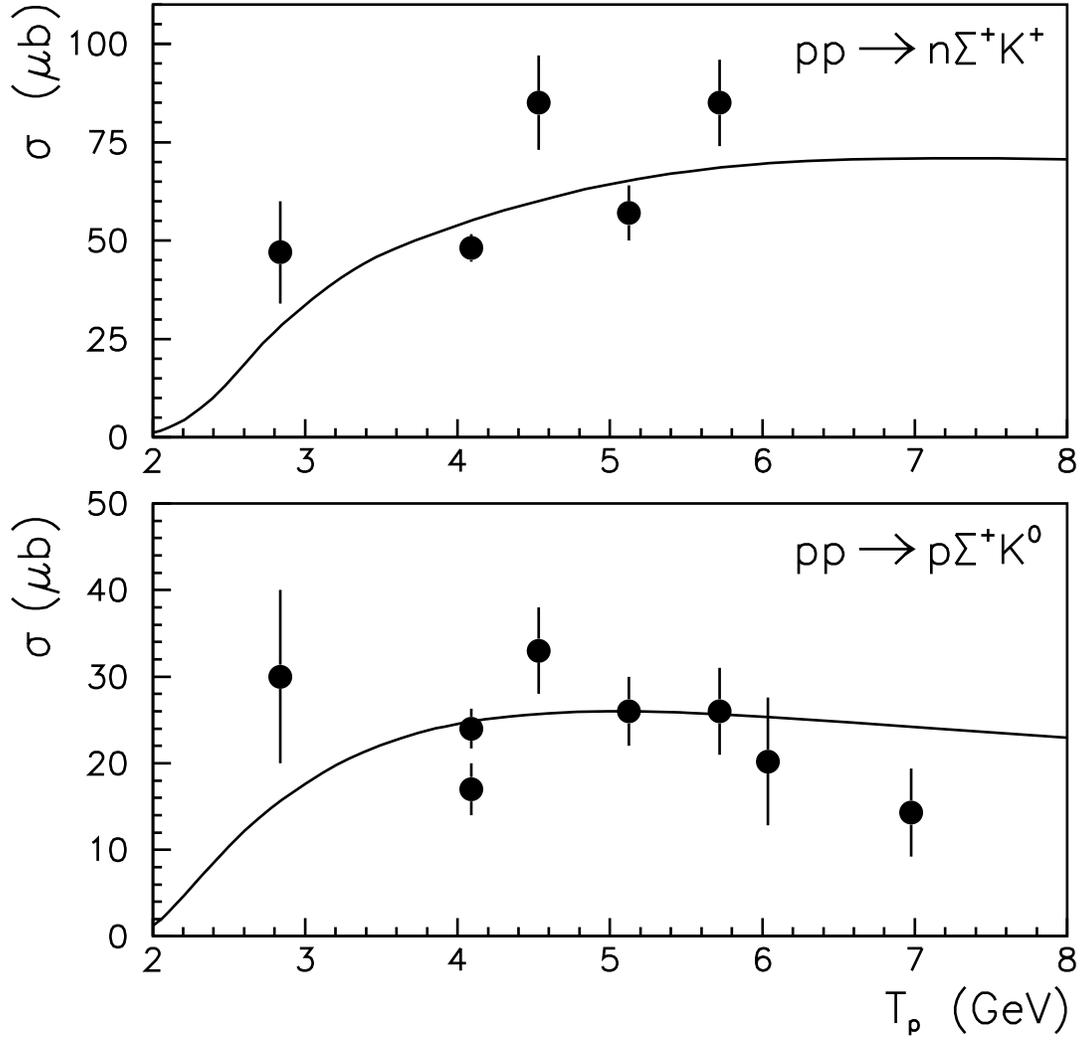,width=16cm}
\caption{\label{ka5}The cross section for the reactions
$pp \to n \Sigma^+ K^+$ and  $pp \to p \Sigma^+ K^0$ in
comparison to the experimental data from~\protect\cite{LB}. 
The lines show the results from the boson exchange model.}
\end{figure}

Within the one-pion exchange model we now also calculate the cross 
sections for the reactions $pp \to p {\Sigma}^+ K^0$,
$pp \to p {\Sigma}^0 K^+$ and
$pp \to n {\Sigma}^+ K^+$ involving a $\Sigma$ in the final channel. 
Our model results are shown in Figs.~\ref{ka5-1},\ref{ka5} 
together with the experimental data.
The calculations quite reasonably reproduce the data in 
case of $K \Sigma$ production and show not much room 
for the contribution from the kaon exchange graph. 

\section{The kaon exchange model}
The kaon exchange diagrams are shown in Fig.~\ref{dia2} and the 
relevant cross sections are similar to~(\ref{OME}) 
\begin{eqnarray}
\label{OKE}
\sigma (pp \to NYK, \sqrt{s}) = \frac {m_N m_Y} {2 {\pi}^2 p_i^2 s}
\int _{W_{min}}^{W_{max}} k \  W^2 \ \sigma (K N \to K N, W) \ dW 
\times \nonumber \\
\int _{t_-}^{t^+} \frac {f^2_{NYK} }{m_K^2} \  F^2(t) 
\ D^2(t) \ t  \ dt ,
\end{eqnarray}
where $m_Y$ is the mass of the produced hyperon ($\Lambda $ or
$\Sigma $) and $W$ is the invariant mass of the 
kaon-nucleon system with
\begin{equation}
W_{min}=m_N +m_K, \ \ \ W_{max}= \sqrt{s}-m_Y .
\end{equation}
Here $t$ is the squared four-momentum
transfer from the initial proton to the hyperon and
\begin{equation}
t_{\pm} = m_N^2+m_Y^2-2E_iE_Y \pm 2q_iq_Y ,
\end{equation}
where $E_i$, $q_i$ are the same as in~(\ref{OME}), while
$E_Y$, $q_Y$ are the energy and three-momentum of the
hyperon in the center-of-mass frame. 
Note that $k$ in this case denotes the  momentum
of the exchange kaon, 
\begin{eqnarray}
k^2 & = & {\lambda}(W^2, m_N^2, m_K^2)/4W^2 \nonumber \\
q_Y^2 & = & {\lambda}(s, W^2, m_Y^2)/4s .
\end{eqnarray}  
In Eq.~\ref{OKE}\ $\sigma (K N \to K N, W)$ is the 
cross section for the $K N \to K N$ reaction.
For $K^+p \to K^+p$ elastic scattering we adopt
the parametrization from Cugnon et al.~\cite{Cugnon2}
\begin{equation}
\sigma (K^+p \to K^+p) =3+11.5 \left[ 1 +
exp \left( \frac {p-1.06} {0.8} \right) \right]^{-1}
\end{equation}
with the cross section given in $mb$ and the laboratory
kaon momentum $p$ in GeV/c. 
Other $KN \to KN$ reaction channels - relevant for 
$\Sigma$-hyperon production (cf. Fig.~\ref{dia2}) - were
taken from Ref.~\cite{Sibirtsev4}; the latter are not 
crucial since the coupling in the $N\Sigma K$ vertex is small.

The kaon propagator was taken similar to the
pion  replacing the $\pi $-meson mass  by the kaon mass.
We use the kaon form factor  
\begin{equation}
F(t) = \frac {{\Lambda}_K^2 - {\mu}^2} { {\Lambda}_K^2 - t}
\end{equation}
with a cut-off parameter ${\Lambda}_K$ that has to be determined
in comparison to the experimental data.
The coupling constant $f^2_{N \Lambda K}$ has to be 
fitted by  the difference between the experimental cross
section for the reaction $pp \to p \Lambda K^+$ and
the results from the pion exchange model. In a similar 
way we get $f^2_{KN \Sigma}$ using the $pp \to p \Sigma K$
reaction channels. The fitting procedure leads to the following
parameters for the kaon exchange model
\begin{equation}
{\Lambda}_K=0.81 \pm 0.14 \ \ GeV , \ \ \ 
f^2_{N \Lambda K}=14.6 \pm 3.1 , \ \ \ 
f^2_{N \Sigma K}=0.88 \pm 0.26 .
\end{equation}

The relation between $f_{NYK}$
and  $g_{NYK}$ coupling constants is
\begin{equation}
\frac{f^2}{4\pi} = \frac{g^2}{4\pi} \ 
\frac{\mu^2}{4m_Nm_Y} 
\end{equation}
were $\mu$ is the kaon mass and $m_Y$ is the  mass of the $\Lambda$ or 
$\Sigma$-hyperon. Therefore the ratio of the $g_{N\Lambda K}$ and 
$g_{N \Sigma K}$ coupling constant equals to $15 \pm 6$,
which is in reasonable agreement with the $SU(6)$ prediction.

\begin{figure}[h]
\epsfig{file=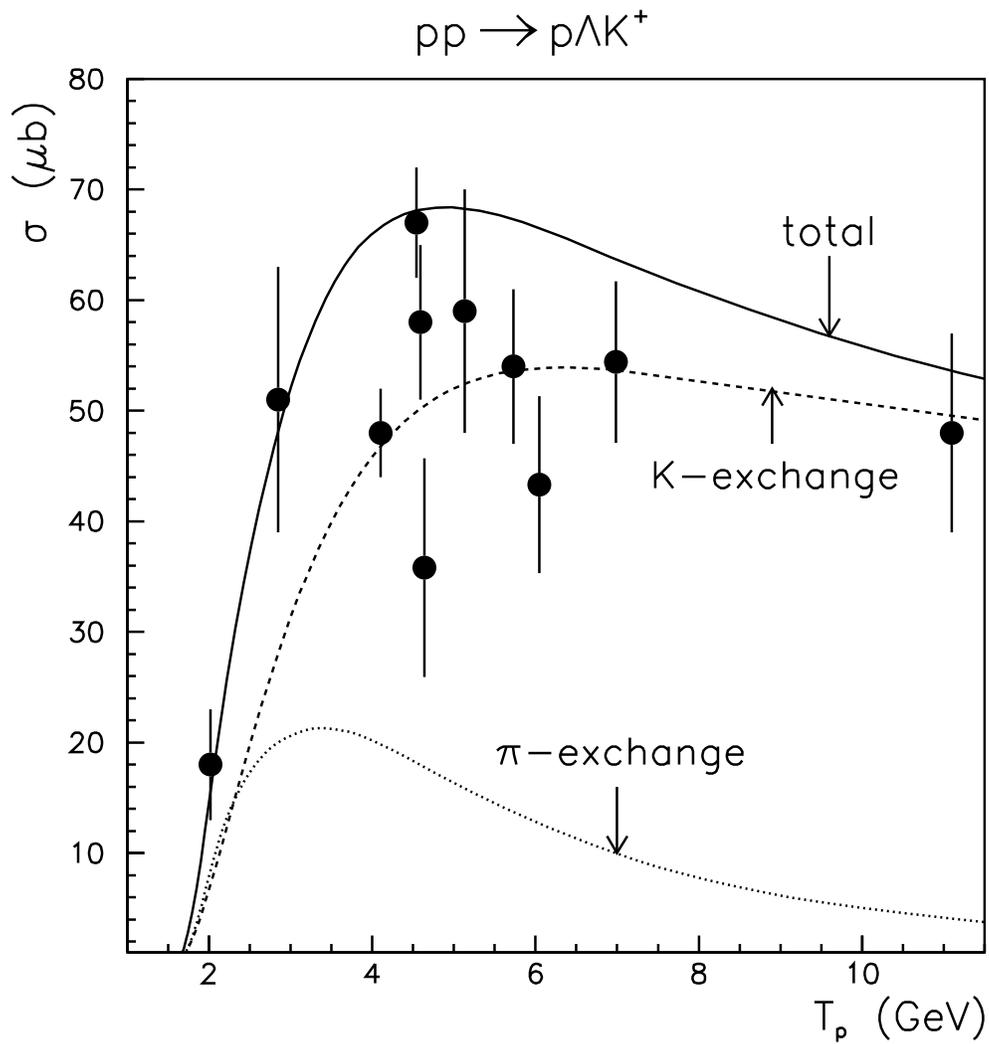,width=15cm}
\caption{\label{ka4}The cross section for the reaction
$pp \to p \Lambda K^+$ in comparison with the experimental 
data from \protect\cite{LB}. The lines show the results from 
the pion (dotted) and the kaon exchange model (dashed),
while the  solid line is the sum of both.}
\end{figure}

In Fig.~\ref{ka4} we display the contribution from 
the kaon exchange graph to the
$pp \to p \Lambda K^+$ cross sections (dashed line)
and the results obtained with the pion exchange
model (dotted line). The sum of both contributions 
is shown by the solid line and perfectly describes the 
data at high as well as low energies. 

Fig.~\ref{ka5-1} show the results calculated for the $\Sigma$-hyperon
production. Unfortunately, the small number of the experimental points
and large experimental errors substantially spoil the analysis.

\section{Near-threshold behavior of the cross section}
We note that different theoretical 
models~\cite{Sibirtsev1,LiKo,Wilkin,Tsushima1} predict a
very similar energy dependence of the $pp \to p\Lambda K^+$
cross section close to the reaction threshold, i.e.
at excess energies 
$\epsilon = \sqrt{s}-m_N-m_{\Lambda}-m_K \le 100$~MeV. Moreover, this
energy dependence reflects essentially the phase space of
the production cross section which is expressed 
through Eq.~(\ref{form4}) by using a constant amplitude for
the reaction as
\begin{equation}
\label{phasa}
\sigma = \frac {R_3}{I} \ |M|^2
\end{equation}
with $R_3$ denoting the three-body phase space integral 
\begin{equation}
R_3 = \frac{\pi^2}{4s} \int_{(m_{\Lambda}+m_K)^2}^{(\sqrt{s}-m_N)^2}
\lambda^{1/2}(s,s_1,m_N^2)
\ \lambda^{1/2}(s_1,m_{\Lambda}^2,m_K^2) 
\ \frac{ds_1}{s_1} ,
\end{equation}
while $I$ is the flux factor~\cite{Byckling}
\begin{equation}
I=2 \ (2\pi)^5 \ \lambda^{1/2}(s,m_N^2,m_N^2).
\end{equation}

\begin{figure}[h]
\epsfig{file=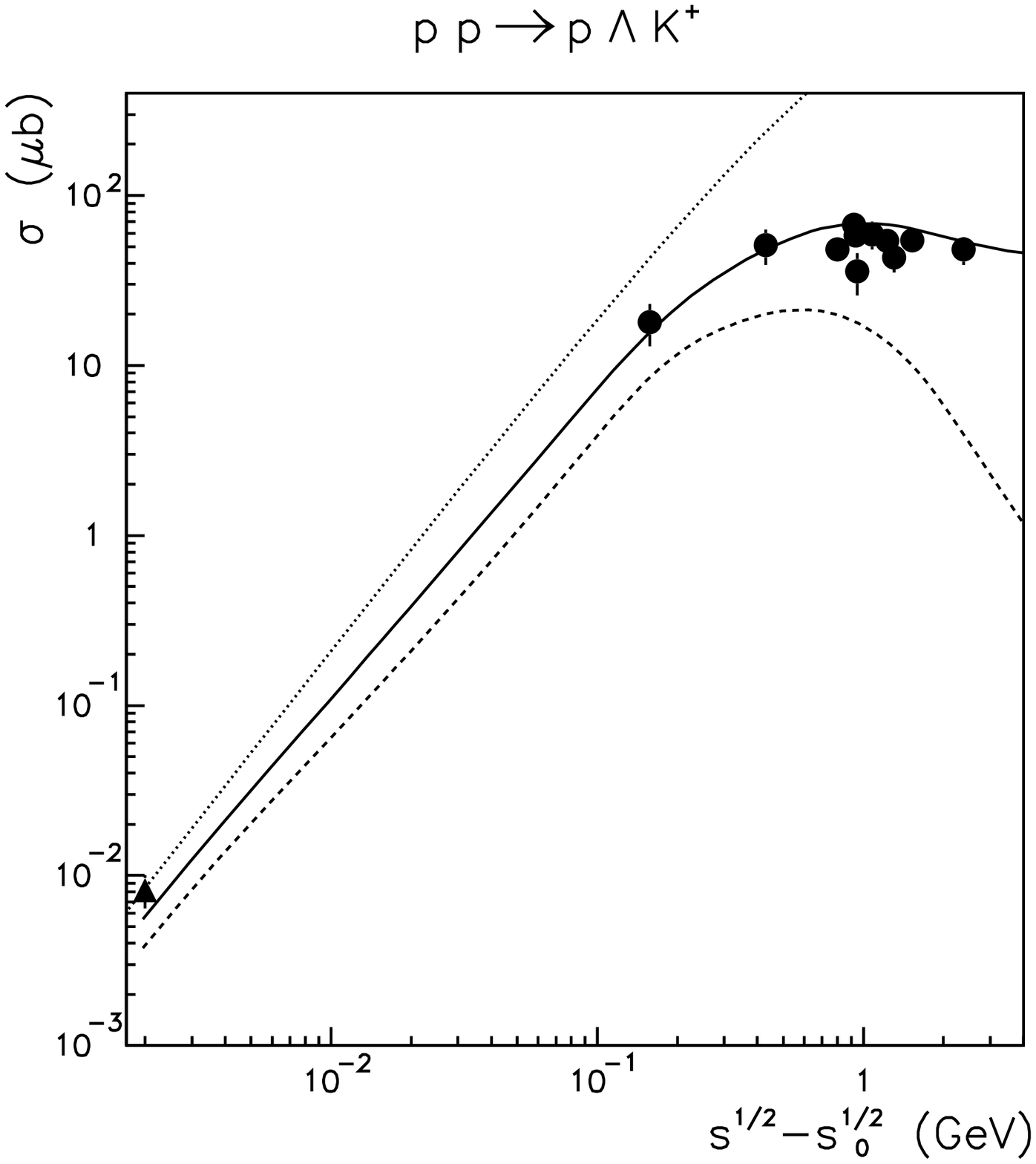,width=15cm}
\caption{\label{cr1}The $pp \to p \Lambda K^+$
cross section as function of the excess energy. 
Dots are the experimental data from~\protect\cite{LB} while
the triangle is from~\protect\cite{Balewski}. The solid line shows our
calculation with both pion and kaon exchanges, while
the dashed indicates the result from one pion exchange only.
The dotted line illustrates the phase space dependence
of the production cross section.}
\end{figure}

The cross section calculated with 
constant amplitude is shown in Fig.~\ref{cr1} together with the
experimental data~\cite{LB,Balewski} as a function of the excess
energy. Here we fit $|M|^2$ in order to match the
experimental point from Ref.~\cite{Balewski}.

Let us now discuss why both our calculations and the
results from the resonance model follow the phase space consideration.  
This similarity can be explained by the kinematical constraints of the
reaction at low $\epsilon$. Note that in any of the models the
integration of the general expression (i.e. Eq.~(\ref{form4})) has
to be performed over the variables $t$ and $s_1$, which depend
only on the collision energy as well as on the 
mass of the final particle.
For a fixed value of the excess energy $\epsilon$ 
the limits of the integration over $s_1$ are defined by
\begin{equation}
m_{\Lambda}+m_K \le \sqrt{s_1} \le m_{\Lambda}+m_K +\epsilon
\end{equation}
for the pion exchange and
\begin{equation}
m_N+m_K \le \sqrt{s_1} \le m_N+m_K +\epsilon
\end{equation}
for the kaon exchange. Thus the range of the integration
is equal to the excess energy independent from the production
mechanism. The relevant amplitude $M_1$ of the corresponding subprocess
(as well as cross section) is almost
energy independent for 
$\epsilon \le 100$~MeV. Moreover, some of the earlier calculations 
on strangeness production~\cite{Yao,Ferrari3} simply neglect this 
energy dependence of the $|M_1|$ amplitude even for higher 
excess energies.

At low $\epsilon$ the  $t$-dependence seems to be more crucial. However,
at the reaction threshold $\sqrt{s}=m_N+m_{\Lambda}+m_K$ 
the momenta of the produced particles in the center-of-mass of 
the colliding nucleons are close to zero and the momentum transfer 
is determined as
\begin{equation}
t=2m_N^2-m_N\sqrt{s}=m_N(m_N-m_{\Lambda}-m_K)\simeq -0.63 \ \ GeV^2 .
\end{equation}

\begin{figure}[h]
\epsfig{file=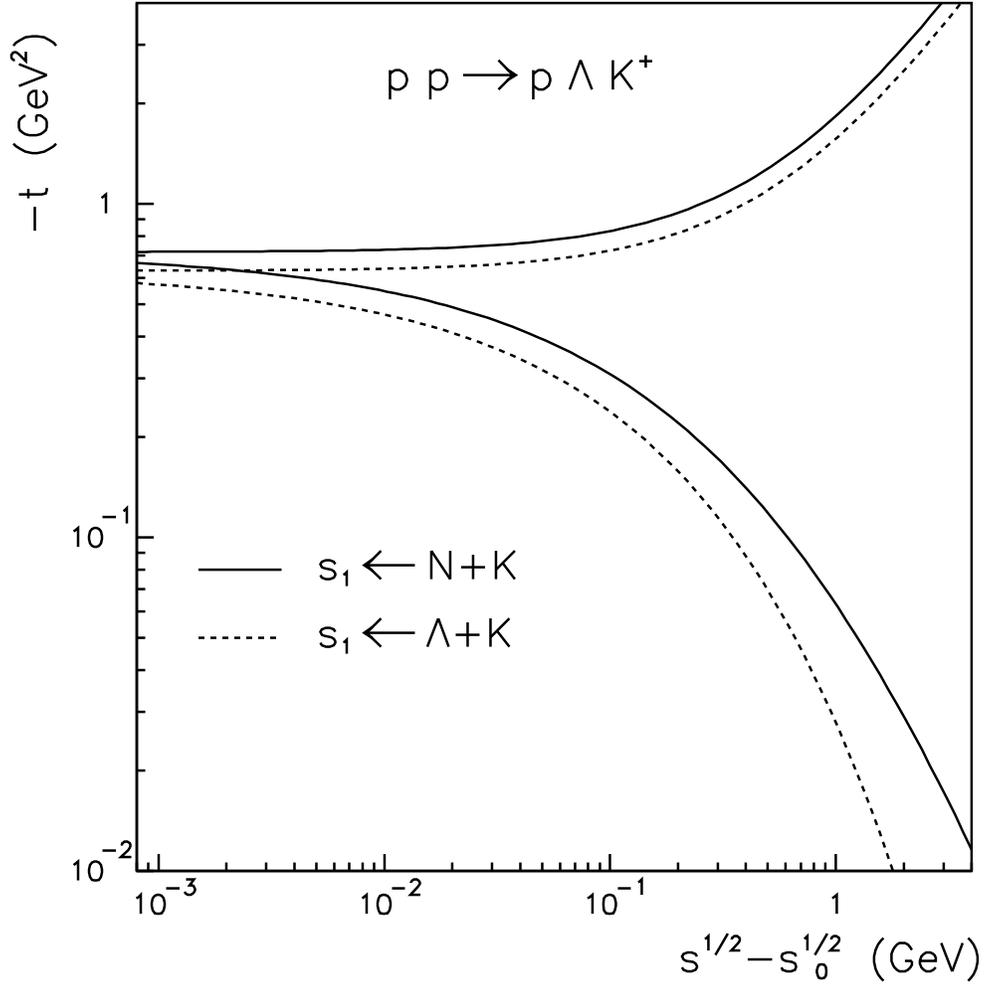,width=15cm}
\caption{\label{cr2}The momentum transfer as a function of the
available energy above the $pp \to p \Lambda K^+$ reaction threshold.
The lines indicate lower and upper limits.  The solid lines stand
for the one kaon exchange while the dashed are for 
the pion exchange model.}
\end{figure}

Furthermore, lets look at the dependence of the
momentum transfer on the collision energy.
Fig.~\ref{cr2} shows the lower and upper limits
of $-t$ as a function of the available energy
$\epsilon$ above the reaction threshold. 
Since the momentum transfer also depends on the
invariant mass of the particle in the upper vertices
of the diagrams shown in Figs.~\ref{dia1},\ref{dia2},
we vary $\sqrt{s_1}$ within the range $\epsilon$
and display only the minimal and maximal values.
Note that within the range
$\sqrt{s}-\sqrt{s_0}<100$~MeV  $t$ changes only slightly and 
is almost constant.

Thus the form factor and the propagator part of the 
reaction amplitude are almost constant as
well as the amplitude itself. Consequently the
production cross section at low $\epsilon$ follows
the phase space dependence according to Eq.~(\ref{phasa}).
Moreover, this property is fundamental for most
of the reactions with particle production near threshold.

Apart from the latter considerations there are several
factors which might change the energy dependence of the
cross section relative to that from phase space. Most effective are
interactions in the initial and final states. Following the
effective range approximation one can assume that 
these interactions depend on the relative momentum of the particles.
Particularly for strangeness production from $pp$
collisions the relative momentum of the colliding nucleons
is quite large in the initial state and their interaction here
is suppressed. This is not the case for the final state where the
strong hyperon-nucleon
interaction can affect the total
cross section close to threshold.

Fig.~\ref{cr1} shows the near-threshold behavior of the
$pp \to p \Lambda K^+$ cross section calculated with
our model and with the parameters fitted by the experimental
data at higher energies. Our results reasonably describe
the experimental point from COSY at 
$\epsilon =2$~MeV~\cite{Balewski}

\section{Comparison with other $pp$ parameterizations}
The calculated cross sections for the reaction $pp \to NYK$
can be parameterized as
\begin{equation}
\label{PARM}
\sigma (pp \to NYK, \sqrt{s})= a 
{\left( 1- \frac {s_0} {s} \right) }^b 
{\left( \frac {s_0} {s} \right) }^c
\end{equation}
with $s_0$=6.487 GeV$^2$ for $\Lambda $ and 6.864 GeV$^2$
for $\Sigma $ production. The parameters $a$, $b$ and $c$ are
obtained by fitting the results from the boson exchange model 
taking into account the contribution from both pion 
and kaon exchanges. In Table 3 we show the parameters for the
different reaction channels.

\begin{table}[h]
\begin{center}
\caption{Parameters of the approximation (\protect\ref{PARM}).}
\vspace{0.5cm}
\begin{tabular}{|l|c|c|c|}
\hline
Reaction & $a$ [$\mu $b] & $b$  & $c$ \\
\hline
$pp \to p \Lambda K^+ $ & 732.16 & 1.8 & 1.5 \\
$pp \to p {\Sigma}^+ K^0 $ & 338.46 & 2.25 & 1.35 \\
$pp \to p {\Sigma}^0 K^+ $ & 275.27 & 1.98 & 1.0 \\
$pp \to n {\Sigma}^+ K^+ $ & 549.51 & 1.87 & 0.98 \\ 
\hline 
\end{tabular}
\end{center}
\end{table}

\begin{figure}[h]
\epsfig{file=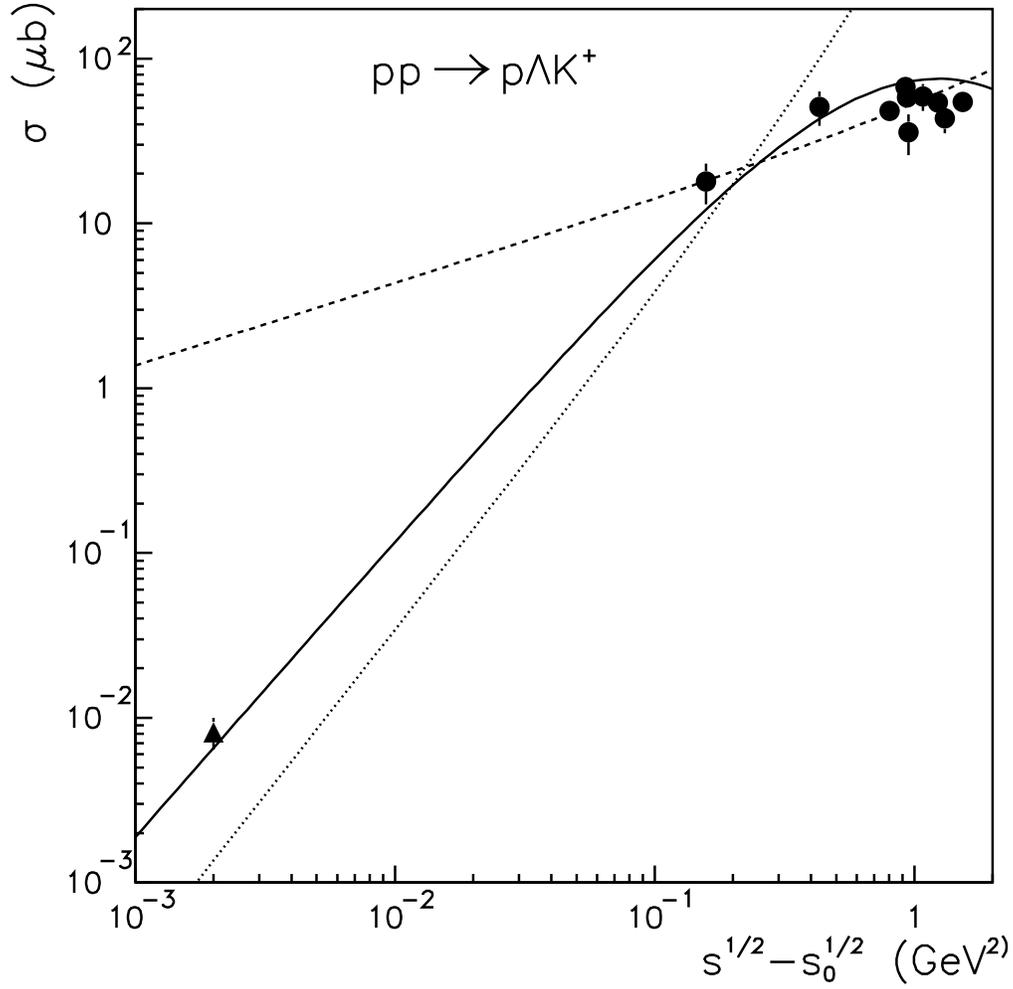,width=15cm}
\caption{\label{ka3}The cross section for the reaction 
$pp \to p \Lambda K^+$. The dots show the experimental data
\protect\cite{LB} while the triangle represents the result from 
COSY~\protect\cite{Balewski}. The solid line shows the 
results from the boson exchange model. The dashed line  is 
the parameterization from Randrup and Ko~(\protect\ref{RAN}) 
while the dotted line indicates the result from
Sch\"urmann and Zwermann~(\protect\ref{SHU}).}
\end{figure}

Our parameterization for the reaction $pp \to p \Lambda K^+$
is shown in Fig.~\ref{ka3} together with the results from 
the boson exchange
model and the experimental data. The dashed line shows the
parameterization proposed by Randrup and Ko~\cite{Randrup1}
\begin{equation}
\label{RAN}
\sigma (pp \to p \Lambda K^+) = 24 \frac {p_{max}} {m_K}
\ \ \ [\mu b]
\end{equation}
with $p_{max}$ given by 
\begin{equation} 
p_{max}^2= \lambda(s, {\left[m_N+m_{\Lambda} \right]}^2, m_K^2)/4s.
\end{equation}
The dotted line in Fig.~\ref{ka3} indicates the parameterization from
Sch\"urmann and Zwermann~\cite{Zwermann}
\begin{equation}
\label{SHU}
\sigma (pp \to  K^+X) = 31.7 { \left( \frac {p_{max}} {m_K}
\right) }^4
\ \ \ [\mu b] .
\end{equation}
Since the  parameterization~\cite{Zwermann} reflects the phase 
space energy dependence of the production  cross section it
is close to the results obtained with the boson exchange
model.

\begin{figure}[h]
\epsfig{file=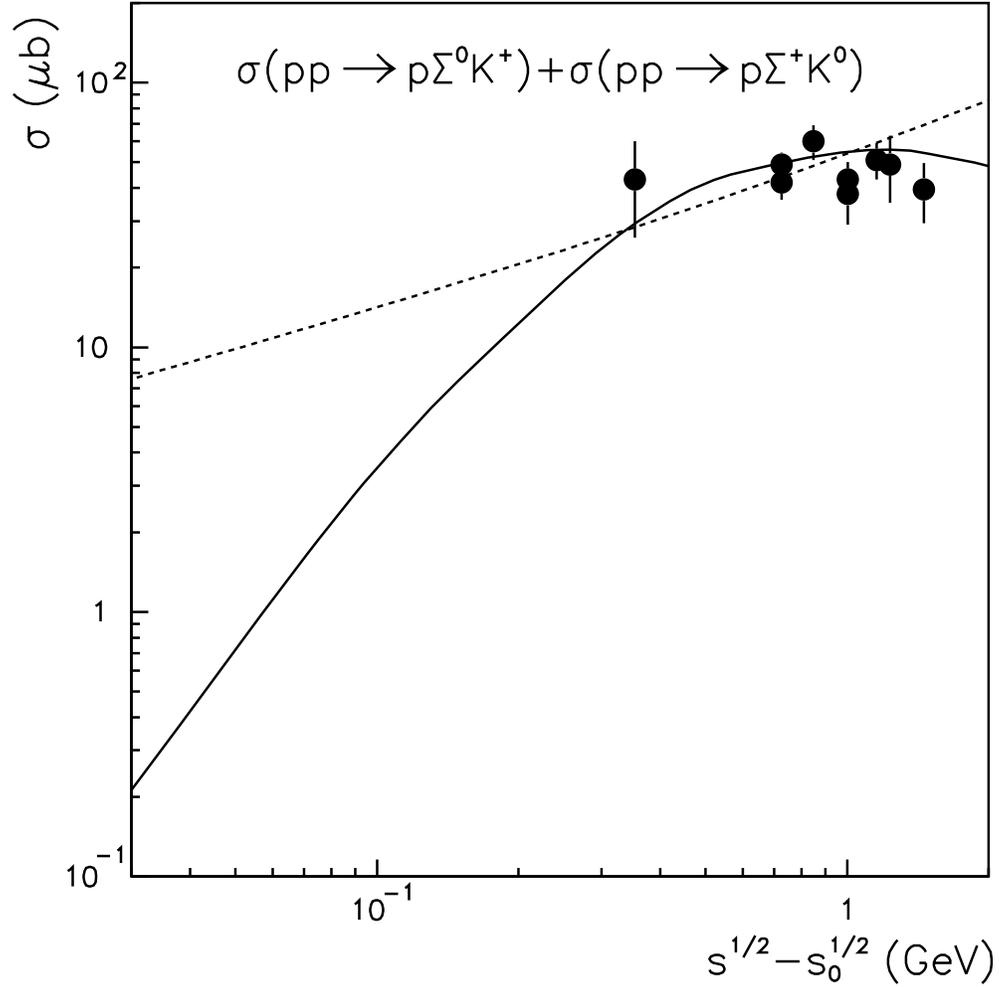,width=15cm}
\caption{\label{ka11}Sum of the 
$pp \to p {\Sigma}^0 K^+$ and $pp \to p {\Sigma}^+ K^0$
cross sections. The dots  show the experimental data
\protect\cite{LB}. The solid line shows the result from
the boson exchange model while the dashed line corresponds to the 
parameterization~(\protect\ref{RAN}).}
\end{figure}

The cross section for the reaction channels with a $\Sigma $ hyperon
is parameterized in~\cite{Randrup1} as
\begin{equation}
\label{SIG}
\sigma (pp \to p {\Sigma}^0 K^+) +
\sigma (pp \to p {\Sigma}^+ K^0) =
24 \frac {p_{max}} {m_K}
\ \ \ [\mu b]
\end{equation}
with the maximal kaon momentum $p_{max}$ for the $N \Sigma K$  
channel  
\begin{equation} 
p_{max}^2= \lambda (s, (m_N+m_{\Sigma})^2, m_K^2).
\end{equation}  
In Fig.~\ref{ka11} we show~(\ref{SIG}) together with our 
result. The cross section calculated within the boson exchange model is 
substantially smaller then the parameterization from  
Ref.~\cite{Randrup1}.

\begin{figure}[h]
\epsfig{file=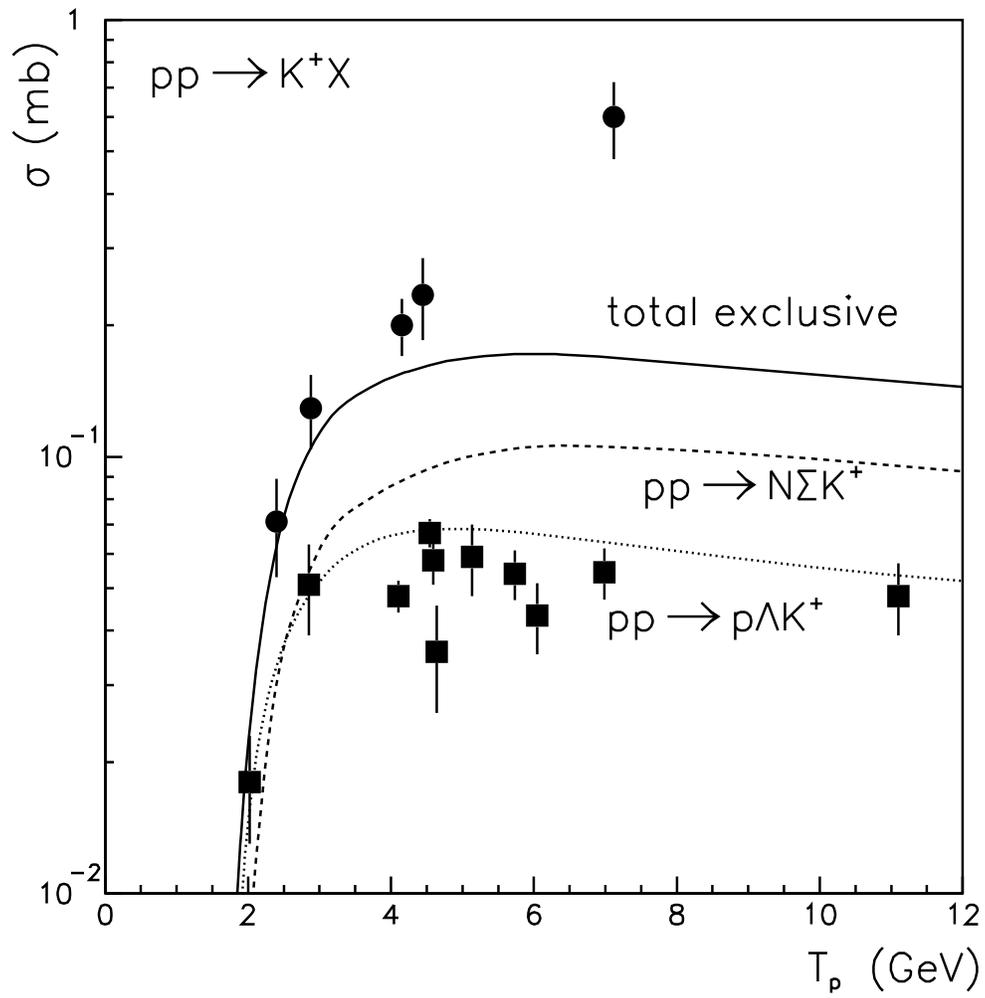,width=15cm}
\caption{\label{ka6}The cross section for inclusive 
$K^+$ production (full dots)
and the reaction  $pp \to p \Lambda K^+$ (squares).
The lines show the contributions from the $\Lambda $ channel -(dotted),
$\Sigma $ -(dashed) and the sum -(solid) calculated according to
(\protect\ref{PARM}).}
\end{figure}

The cross section for the inclusive $K^+$-meson production from
$pp$ collisions is shown in Fig.~\ref{ka6} as a function of 
the bombarding  energy. 
The squares show the experimental data for the 
reaction $pp \to p \Lambda K^+$ while the dotted line is
our parameterization~(\ref{PARM}) for this channel. The dots
are the experimental cross section for the inclusive reaction
$pp \to K^+ X$. The dashed line shows the contribution from 
$pp \to N \Sigma K^+$ while the solid line is the
sum of the $\Lambda $ and $\Sigma $ channels. 
Note that at
incident proton energies above 2.5 GeV the contributions from the
$\Sigma $ reaction channels become larger than those from the $\Lambda $
channel, which should be important for transport simulations of the
$K^+$ production. At bombarding energies above
3.5 GeV the calculated cross section of the reaction
$pp \to N Y K^+ $ underestimates the
experimental data for inclusive kaon production 
substantially which demonstrates
the importance of final channels with additional $\pi $-mesons.

\section{Summary}
Within the boson exchange model we have calculated 
the cross section for the
reactions $pp \to N Y K$ with a $\Lambda $, $\Sigma $-hyperon
and $K^0$, $K^+$-meson. The contributions from the one-pion 
and kaon exchange were studied separately. 
We find that the dominant contribution to the reaction
$pp \to p \Lambda K^+$ stems from the kaon exchange diagram,
whereas $\Sigma $-hyperon  production is dominated by one-pion exchange.
The boson exchange model rather well describes the available
experimental data on the reaction $pp \to N Y K$.
Furthermore, 
the ratio of the $N\Lambda K$ and $N\Sigma K$ coupling constants
fitted to the experimental data is in good agreement with
the $SU(6)$ prediction.

The near-threshold behavior of the production cross section
was analyzed and shown to be dominated by phase space. 
The calculated cross sections were parameterized and compared
with other parameterizations that have been widely used in
proton-nucleus and heavy-ion simulations before.

It is found that the $pp\to N\Sigma K^+$ reaction dominates 
over $\Lambda$-hyperon production with
increasing beam energy.
The inclusive $K^+$-meson production is described by the sum of the
three body final state channels up to the incident proton energy of
3.5 GeV. At higher energies  additional $\pi $-mesons in 
the final channel have to be taken into  account.

\vspace{1cm}
The authors gratefully acknowledge stimulating discussions with
L.~Jarczyk, B.~Kamys, C.M.~Ko, U.~Mosel, Z.~Rudy,
A.~Strzalkowski and  K.~Tsushima.
Furthermore, the authors like to thank the Institute
of Physics at the Jagellonian University of Cracow for
warm hospitality during their research visits.
 
\newpage

\newpage

\newpage

\newpage

\newpage

\newpage

\newpage

\newpage

\newpage

\newpage

\newpage

\newpage

\newpage

\newpage

\end{document}